\documentclass[reprint,amsmath,amssymb,aps,prl]{revtex4-1}
\usepackage[dvips]{graphicx}
\usepackage{dcolumn}
\usepackage{bm}
\usepackage{hyperref}
\usepackage{color}
\usepackage{cancel}
\usepackage{bmpsize}
\usepackage{amsmath}
\usepackage{amssymb}
\usepackage{epstopdf}

\newcommand{\avg}[1]{\left\langle #1\right\rangle}
\newcommand{\be}[0]{\begin{equation}}
\newcommand{\ee}[0]{\end{equation}}
\newcommand{\intinf}[0]{\int\limits_{-\infty}^{+\infty}}
\newcommand{\iintinf}[0]{\iint\limits_{-\infty}^{+\infty}}
\newcommand{\iiintinf}[0]{\iiint\limits_{-\infty}^{+\infty}}

\newcommand{\lra}\simeq
\newcommand{\eeqref}[1]{Eq.~(\ref{#1})}

\newcommand{\de}[0] {{\rm d}}

\begin{document}

\title{Far-field linear optical superresolution \\via heterodyne detection in a higher-order local oscillator mode}

%\iffalse

\author{Fan Yang$^1$, Arina Taschilina$^1$, E. S. Moiseev$^1$, Christoph Simon$^1$, A. I. Lvovsky$^{1,2}$}\email{trekut@gmail.com}
\affiliation{$^1$Department of Physics and Astronomy, University of Calgary, Calgary, Canada, T2N 1N4}
\affiliation{$^2$Russian Quantum Center, 100 Novaya St., Skolkovo,
	Moscow region, 143025, Russia}

%\affiliation{$^4$Institute for Quantum Science and Technology, University of Calgary, Calgary AB T2N 1N4, Canada}

%Include the email address of the corresponding author here.
%\email{LVOV@ucalgary.ca}
%\date{\today}
%\fi
%
\begin{abstract}
%Traditional optical microscopy is limited in resolution by the Rayleigh criterion. 
The Rayleigh limit has so far applied to all microscopy techniques that rely on linear optical interaction and detection in the far field. Here we demonstrate that detecting the light emitted by an object in higher-order transverse electromagnetic modes (TEMs) can help achieving sub-Rayleigh precision for a variety of microscopy-related tasks. Using optical heterodyne detection in TEM$_{01}$, we measure the position of coherently and incoherently emitting objects to within 0.0015 and 0.012 of the Rayleigh limit, respectively, and determine the distance between two incoherently emitting objects positioned within 0.28 of the Rayleigh limit with a precision of 0.019 of the  Rayleigh limit. Heterodyne detection in multiple higher-order TEMs enables full imaging with resolution significantly below the Rayleigh limit in a way that is reminiscent of quantum tomography of optical states.
\end{abstract}

\maketitle
\vspace{10 mm}

\paragraph{Introduction.}
Since the invention of the optical microscope, there has been a quest for enhancing its resolution. The Rayleigh criterion \cite{Hecht} establishes the minimum resolvable distance in a direct image of a pair of sources to be limited by diffraction according to $d_R=1.22\lambda/2\textrm{NA}$, where $\lambda$ is the wavelength and $\textrm{NA}$ is the numerical aperture of the objective lens. In the past century, a number of techniques for circumventing the Rayleigh limit have emerged \cite{HellReview}. These methods rely, for example, on using shorter-wavelength radiation \cite{Xray}, near-field probing \cite{NearField}, nonclassical \cite{QMetro} or nonlinear \cite{STED} optical properties of the object or switching the object's emission on and off \cite{PALM,STORM,GreenSwitch}. However, these approaches are often expensive and not universally applicable. Therefore finding a linear-optical microscopy technique that is operational in the far-field regime remains an important outstanding problem. 

A promising approach to addressing this problem is by detecting the light  emitted by the object in higher-order transverse electromagnetic modes (TEMs). A point source emits primarily into the fundamental TEM$_{00}$ mode. However, when the emitter is displaced from the center of the fundamental mode, higher-order TEMs are illuminated. In this way, null measurements of small displacements are possible. For example, homodyne detection in TEM$_{01}$ has been used  to detect small displacements of a laser beam \cite{Hsu04} and to tracking particles with a nanometer resolution \cite{Taylor13}. These experiments were performed for single emitters only.

More recently, Tsang {\it et al.} showed theoretically that sub-Rayleigh distances between two identical point sources can be estimated by measuring the photon count rate in TEM$_{01}$ \cite{Tsang,TsangSLIVER}. Remarkably, for distances below the Rayleigh limit, the per-photon uncertainty of this measurement is much less than that associated with direct imaging. This feature is particularly valuable when the number of photons the object can radiate is limited, such as in the case of photobleaching.
%In order to measure the intensity of the TEMs, they need to be separated from each other. Reference \cite{Tsang} proposes to solve this task by means of waveguides with sophisticated spatial structures, which is difficult to realize in practice. 
%In a later publication \cite{TsangSLIVER}, Tsang and co-workers propose to sort TEMs by their parity via wavefront inversion. This technique is practically easier and was recently implemented \cite{Ling}. However, this method is less general because it does not permit separation of all transverse modes from each other. 

Inspired by these developments, we overcome the Rayleigh limit by means of heterodyne detection, taking advantage of the fact that the heterodyne detector is only sensitive to the electromagnetic field in the mode of the local oscillator (LO) \cite{RaymerRMP}. %The LO can readily be prepared in a desired TEM by means of a spatial light modulator or, as in our case, a Fabry-Perot cavity. 
 % and permits measuring both the intensity and phase of the light in the given mode. 
We apply the technique in a variety of settings. First, we determine the positions of single objects emitting coherent and incoherent light. Second, we determine the distance between two identical incoherent objects separated by a distance below the Rayleigh limit. In both cases, we demonstrate a measurement precision significantly below that limit. 

In addition, we utilize a mathematical analogy between decomposing an image into TEMs and representing the quantum state of a harmonic oscillator in the Fock basis to propose a new microscopy technique. By measuring intensity and phase of the object's emission into all TEMs one can completely reconstruct its image, in principle with an arbitrarily high precision. In an experiment, the imaging resolution depends on the number of TEMs in which the detection can be realized. %We calculate the imaging resolution as a function of that number.

\paragraph{Concept.} Consider an optical microscope objective lens that is used to image a plane object with transverse spatial field distribution $E(x)$. The field distribution in the image plane is then given by the convolution 
\begin{equation}\label{Eprimeconv}
E'(x')=\intinf E(x) T(x'-x) \de x,
\end{equation}
where $T(\cdot)$ is the transfer function of the objective lens and we have assumed the magnification to be unity for convenience (\emph{Supplementary}). The transfer function can be approximated by a Gaussian 
\begin{equation}\label{TGauss}T(x)\approx \frac1{(2\pi)^{1/4}\sqrt\sigma} e^{-x^2/4\sigma^2},
\end{equation} with the width $\sigma\approx0.21\lambda/{\rm NA}$ \cite{Zhang07}. The narrower the aperture in the lens, the wider the transfer function, and the stronger the distortion of the image.

The heterodyne detector generates a current that is proportional to the overlap between the LO and the signal field
 \begin{equation}\label{Pcoh}
 J=\intinf E'(x')E_{LO}(x')\de x',
 \end{equation}
 where $E_{LO}(x')$ is the spatial profile of the LO. The LO is prepared in the TEM$_{01}$ mode such that the corresponding fundamental TEM$_{00}$ mode  is matched to the image of a point source located at $x=0$:  $E_{LO}(x')=\frac1{(2\pi)^{1/4}\sigma^{3/2}}x'e^{-x'^2/4\sigma^2}$. 
 
For this conceptual discussion, we assume that the source is a point located at position $x_p$, so that $E(x)=\delta(x-x_p)$, in which case we have 
 \begin{equation}\label{Pcoh1}
 J(x_p)=\intinf T(x'-x_p)E_{LO}(x')\de x',
 \end{equation}
which for a Gaussian transfer function reduces to $J(x_p)=\frac 1{2\sigma}x_p e^{-x_p^2/8\sigma^2}$ and the corresponding electronic power
\begin{equation}\label{Pxp}
P(x_p)\propto J^2(x_p)=\frac 1{4\sigma^2}x_p^2e^{-x_p^2/4\sigma^2}. 
\end{equation}
The signal vanishes at $x_p=0$, enabling sensitive null measurement of the source position with respect to the central point \cite{Hsu04}.

Especially useful is the enhancement associated with determining the distance $d$ between two incoherent point sources.  Suppose the sources are located at $x=\pm d/2$, where $d$ is the distance between them. The signal in TEM$_{01}$ is given by $P(d/2)+P(-d/2)$, which is proportional to $d^2$ in the leading order. The intensity in the direct image of this object, on the other hand, is given by $I(x')=S(x'+d/2)+S(x'-d/2)$, where  $S(\cdot)=|T(\cdot)|^2$ is the point spread function of the microscope. Approximating $S(x'\pm d/2)=S(x')\pm\frac d2\frac{\partial}{\partial x}S(x')+O(d^2)$, we find $I(x')=2S(x')+O(d^2)$. The effect on the image due to the separation of the slits is also of the second order in $d$, but  with a macroscopic zeroth-order background. Any noise in this background will have a deleterious effect on the measurement precision of the term of interest. 

\begin{figure}[h]
	\includegraphics[width=\columnwidth]{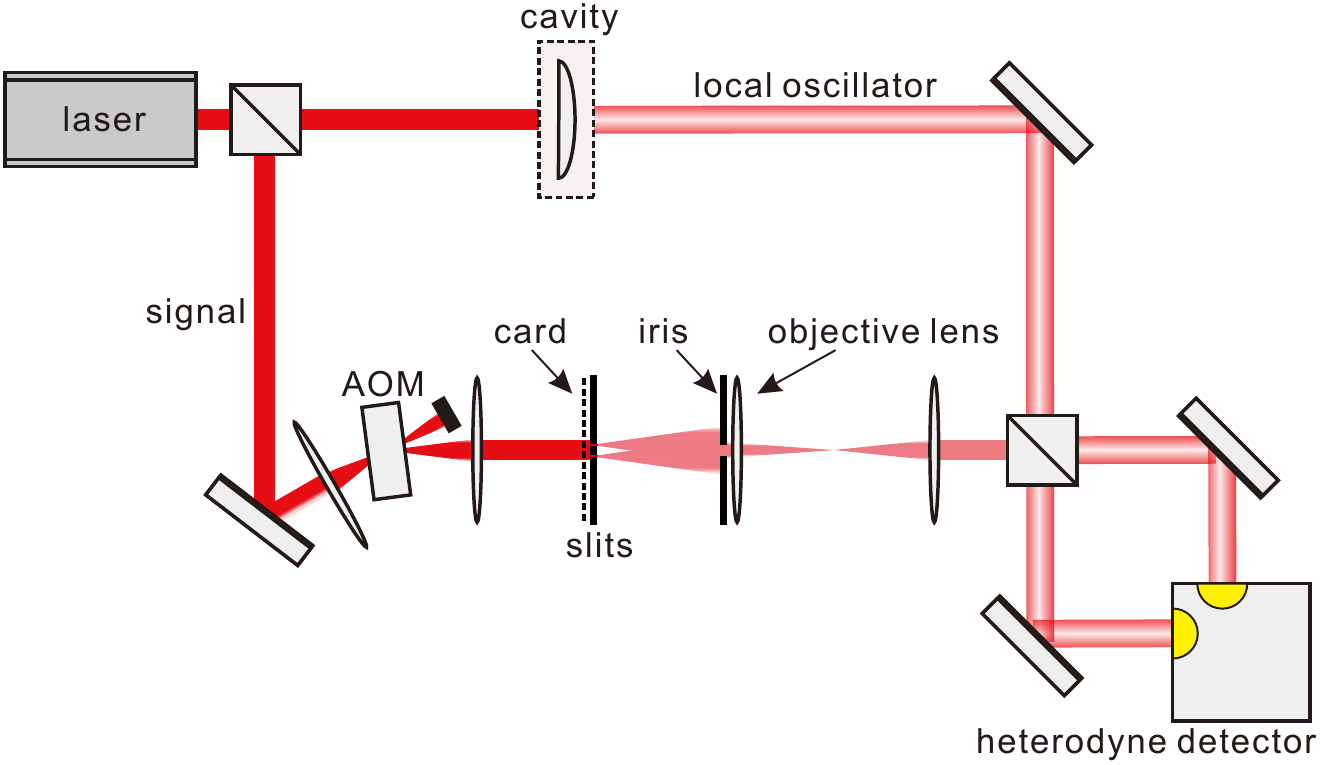}
	\caption{Experimental setup. AOM: acousto-optical modulator. \label{expFig}}
\end{figure}

\paragraph{Experiment.} The experimental setup is shown in Fig.~\ref{expFig}. Both the signal and LO are obtained from a home-made external cavity diode laser, producing $\sim45$ mW at a wavelength of $\lambda=780$ nm. We prepare the LO in the desired TEM by transmitting it through a temperature stabilized monolithic cavity with a finesse of about 275 \cite{pantita}. The laser frequency is locked to the cavity resonance by means of the Pound-Drever-Hall technique \cite{PDH}. The signal beam passes through an acousto-optic modulator operating at 40 MHz in order to reduce the effect of flicker noise in balanced detection. The modulated beam is collimated to an about 5 mm diameter and sent to a diaphragm with four pairs of slits of 0.15 mm width whose centers are separated by $d=0.25$, $0.50$, $0.75$ and $1.00$ mm (3B Scientific U14101). The power of the beam transmitted through the slits is $\sim 200$ $\mu$W.For the measurements with incoherent light, we place a white paper library card before the slits. During the data acquisition, the card is moved in the transverse plane by a motorized translation stage to achieve averaging over the incoherent light statistics. The incoherent optical power behind the slits is $\sim 10$ $\mu$W. After the slits, the light propagates in free space for $L=84$ cm and passes through an iris diaphragm. The diameter of the diaphragm is measured using an optical microscope as $0.8\pm0.1$ mm  and independently estimated with a higher precision from the fit to the data as $0.87\pm0.01$ mm. This latter value corresponds to a numerical aperture of NA$=0.52\times10^{-3}$ and the Rayleigh distance of $d_R=0.912$ mm. 

The field transmitted through the diaphragm is refocused by an objective lens and subjected to heterodyne detection by means of a balanced detector (ThorLabs PDB150A-SP). To align the detector, we first match the signal's mode  to the LO prepared in TEM$_{00}$. Subsequently, the monolithic cavity temperature is changed to transmit the TEM$_{01}$ mode and the LO  is adjusted slightly to minimize the interference with the signal. The output photocurrent from that detector is observed with a spectrum analyzer set to a zero span mode at 40 MHz with a resolution bandwidth of 1 kHz and a video bandwidth of 10 kHz. An average of 100 traces is acquired for each measurement.

\begin{figure}[h]
	\includegraphics[width=\columnwidth]{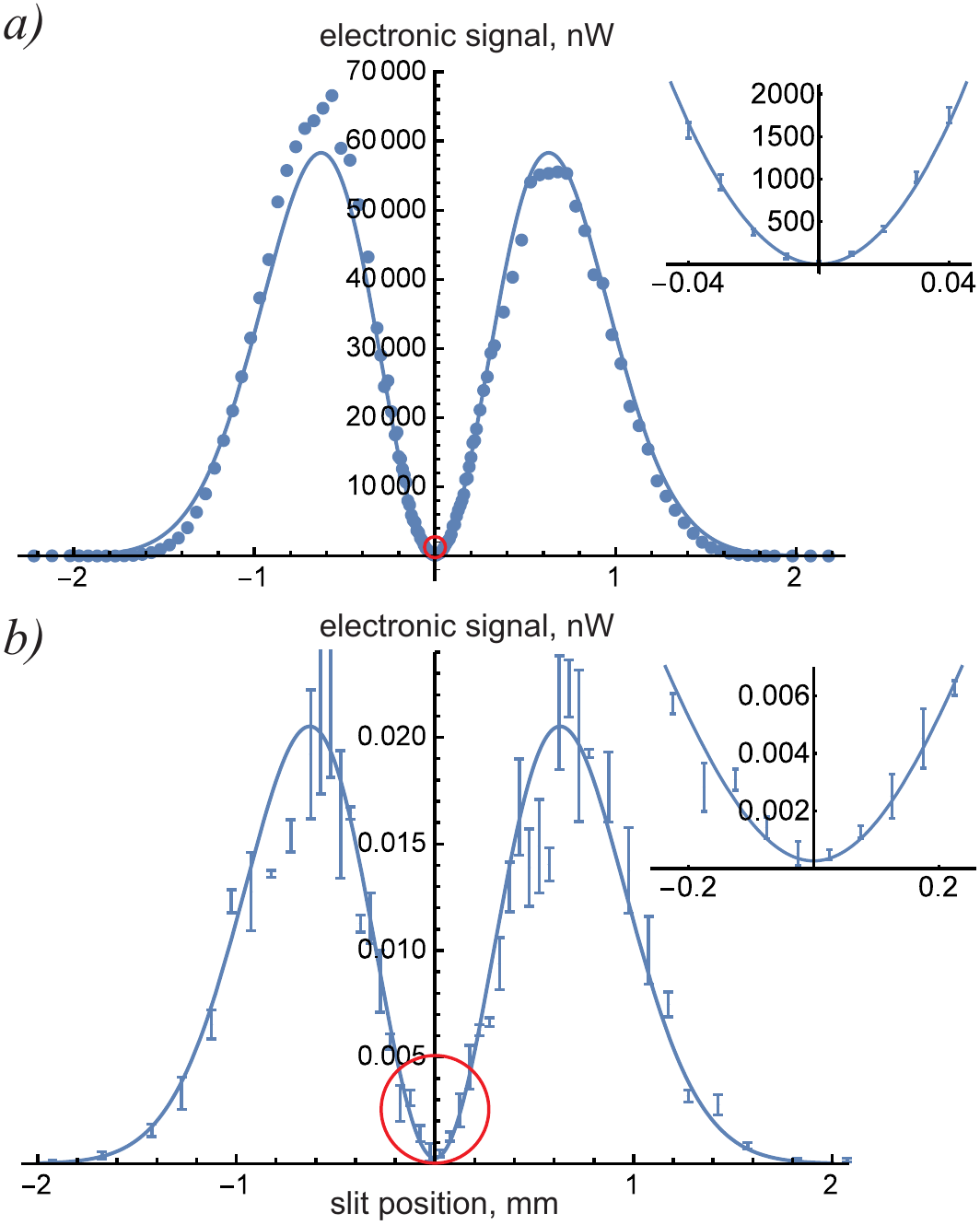}
	\caption{The single slit positioning experiment. The dependence of the heterodyne detector output signal on the slit position measured a) with coherent light, b) with incoherent light is displayed. The theoretical predictions are obtained taking into account the finite width of the slit, but are largely similar to those of \eeqref{Pxp} (\emph{Supplementary}). The theory is fit to the  experimental data by varying the vertical scale and diaphragm diameter. The insets show the areas around the origin, approximately corresponding to the red circles in the corresponding main plots, magnified. \label{singleslitFig}}
\end{figure}

\begin{figure}[h]
	\includegraphics[width=\columnwidth]{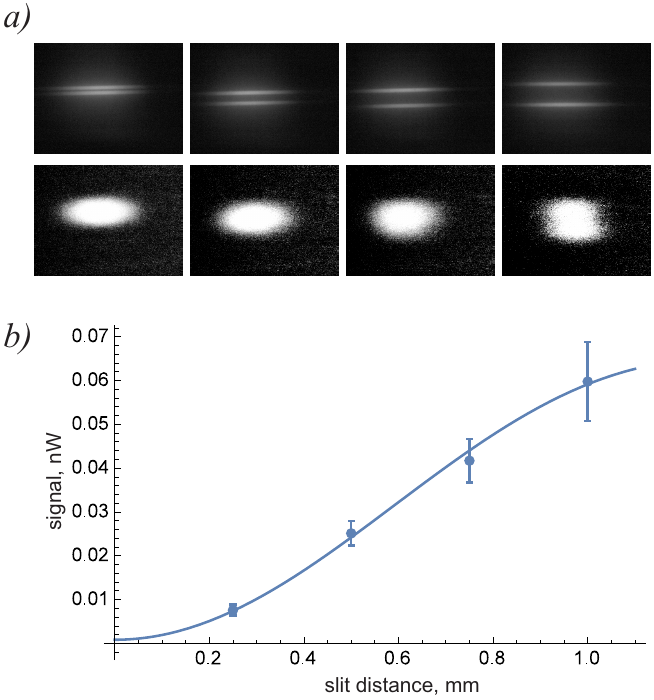}
	\caption{The incoherent double slit experiment. a) Images of the slits with the iris diaphragm fully open (top) and closed to a 0.8 mm diameter (bottom) for $d=0.25,0.50,0.75,1.00$ mm, left to right. The slits with $d<1.00$ mm are not resolved with the closed diaphragm setting. b) Dependence of the signal in TEM$_{01}$ on the slit distance. The error bars show the statistical errors of 12 measurements each. \label{incdoubleFig}}
\end{figure}

\paragraph{Results and discussion.} We first measure the position of a single light source. In this case, the beam only passes through a single 0.15-mm slit, whose position $x_p$ is controlled by a translation stage. We conduct the measurement for both coherent and incoherent light. 

In the coherent case, our setting resembles that of Hsu {\it et al.} \cite{Hsu04}. We acquire the data for each point in Fig.~\ref{singleslitFig}(a) once, except for a set of points around $x_p=0$ shown in the inset. For these latter points, the signal value is acquired ten times to estimate the stochastic experimental error. Two types of experimental imperfections contribute to that error. First, the orthogonality between the LO and signal modes is imperfect and fluctuates due to air movements and  vibration of the optics. As a result, the magnitude of the signal power fluctuation at $x_p=0$ is comparable to the signal itself. Second, the power of the LO  fluctuates due to the instability of the lock of the laser to the cavity resonance. These fluctuations translate directly into those of the detector's output, so the errors at high signals are proportional to the signal. 

In view of this analysis, we model the error of our measurement to behave as $\sigma^2=c^2+(gP)^2$, where $c$ is the uncertainty due to the mode mismatch fluctuations and $bP$ is due to the fluctuations of the LO. The actual experimental  behavior of the error is consistent with this model. %(\emph{Supplementary}). 

We use this model to find the uncertainty of estimating the slit position $x_p$ from the signal power. To this end, we calculate the Fisher information as a function of $x_p$ and find that this function is maximized at $x_p=0.012$ mm. The value of the Fisher information at this point corresponds to the Cram\'er-Rao uncertainty bound \cite{ParamEst} of $\delta x_p=1.4$ $\mu$m, almost three orders of magnitude below the Rayleigh limit (\emph{Supplementary}).

In the experiment with incoherent light [Fig.~\ref{singleslitFig}(b)], the signal's spatial structure is a speckle pattern, changing in time as we move the library card. The intensity of the signal field in the mode being detected is then governed by thermal statistics, so large fluctuations, whose magnitude is on the scale of the signal, are present even for large signals. While the effect of these fluctuations is reduced due to averaging, it is still the dominant source of error for high signal powers. For low signal powers, similarly to the coherent case, the contribution to the uncertainty associated with non-constant mode matching between the signal and LO becomes significant. The general behavior of the experimental error is therefore still consistent with  $\sigma^2=c^2+(gP)^2$, albeit with different values of $c$ and $g$ compared to the coherent case. Experimentally, we evaluate these coefficients by acquiring the heterodyne signal at each $x_p$ three times. We find that the Fisher information is maximized and the uncertainty of $x_p$ is minimized for $x_p=0.14$ mm, corresponding to $\delta x_p=11$ $\mu$m.

Next, we measure the distance between two incoherent light sources. Each slit pair is centered on the laser beam and the output signal is sampled twelve times to estimate the error. The resulting data are shown in Fig.~\ref{incdoubleFig}(b). The error analysis for this setting is similar to that for a single incoherent slit and yields a minimum of $\delta d=17$ $\mu$m for $d=0.18$ mm.

It is interesting to compare the precision of our measurement with what can be achieved by conventional imaging. The images of pairs of slits acquired with a conventional CCD camera are shown in Fig.~\ref{incdoubleFig}(a), bottom row. For separations significantly below the Rayleigh limit, these images do not resolve the slits and cannot be used to determine the separations. Our technique, on the other hand, permits this determination with the precision comparable to the camera pixel size (7 $\mu$m).

At a more quantitative level, the above error analysis implies that the Fisher information decreases to zero, and the estimation uncertainty tends to infinity, for $d\to0$. This is a common feature of both our technique and conventional imaging, arising because of a nonvanishing uncertainty in evaluating the signal  at $d=0$ (\emph{Supplementary}). The fundamental reason for this feature in both cases is the shot noise. In practice, the limitation in our experiment is the fluctuation of the mode matching between the LO and the signal, which can be reduced by using more stable optics. Furthermore, the shot noise level can be reduced by using squeezing \cite{Hsu04,Taylor13}. The conventional imaging technique, on the other hand, is often limited by technical fluctuations of the electronic signal produced by individual pixels in the camera, which complicates precise measurement of minute image width variations associated with varying $d$. Note that the techniques of Refs.~\cite{Tsang,TsangSLIVER} utilize photon counting in higher order modes and do not suffer from the fundamental precision limitation associated with the shot noise. These methods are however still vulnerable to practical noise sources such as detector dark counts.

%While heterodyne detection, in contrast to single-photon measurements in TEM$_{01}$, does not yield quantum advantage in terms of per-photon Fisher information \cite{Tsang}, measurements of small displacements and separations by means of conventional imaging are often obscured by technical noise and the discrete nature of raster images. The pixel size of the camera used to acquire the images, for example, corresponds to about 7 $\mu$m in the image plane, making it a challenge to implement position measurements with a higher precision. The positioning precision that we demonstrated here for the coherent case easily surpasses this limitation.

\begin{figure}[b]
	\includegraphics[width=\columnwidth]{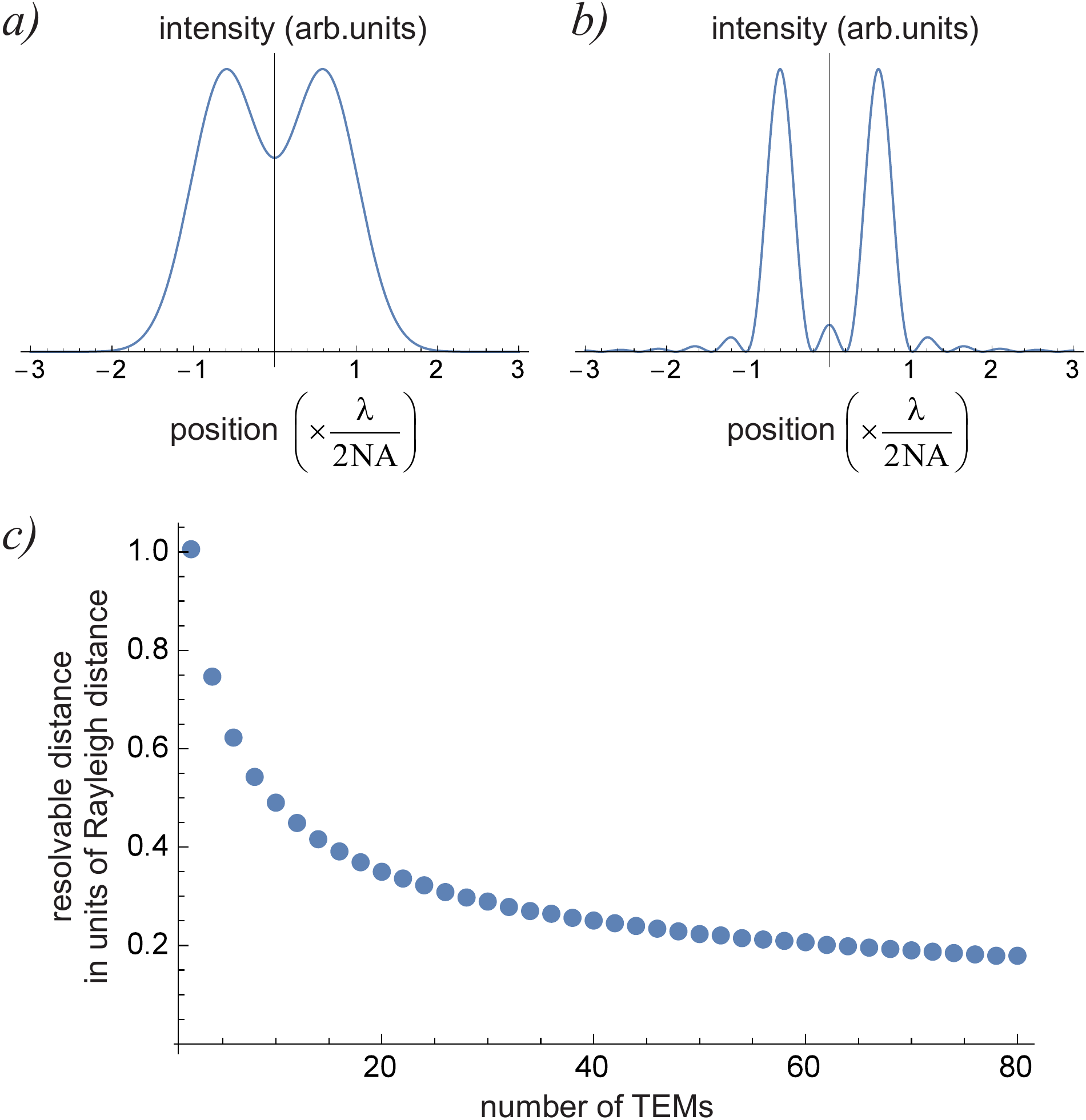}
	\caption{ \label{microscopyFig}Hermite-Gaussian microscopy. a) An image of two single sources positioned at $1.22\lambda/2{\rm NA}$, which corresponds to the Rayleigh limit, expected in conventional imaging. b) An image of the same object expected in HGM with TEM$_{0n}$ for $0\le n\le 20$ exhibits triple enhancement of resolution. c) Resolution of HGM, in units of the Rayleigh distance $1.22\lambda/2\textrm{NA}$, as a function of the number of TEMs used. The resolution is defined as the minimum distance between the point objects such that the image intensity at the center does not exceed 75\% of the maximum intensity.}
\end{figure}

\paragraph{Application to imaging.} Performing heterodyne detection in higher-order Hermite-Gaussian modes TEM$_{0n}$ permits reconstruction of the full image of the object with sub-Rayleigh resolution. To see this, we write the heterodyne detector output photocurrent \eqref{Pcoh} as 
\begin{equation}\label{Imicr}
J_{0n}=\intinf E(x)J(x)\de x,
\end{equation}
where $J(x)$ is the photocurrent in response to a point source at $x$ given by \eeqref{Pcoh1}.
We assume, as previously, that the transfer function is given by \eeqref{TGauss} while the LO is in TEM$_{0n}$ of width $\sigma$: 
\begin{equation}\label{HermiteH}
E_{LO,n}(x)=\frac{H_n(x/\sqrt 2\sigma)}{(2\pi)^{1/4}\sqrt{2^{n}n!\sigma}}e^{-x^2/4\sigma^2},
\end{equation} where $H_n(\cdot)$ are the Hermite polynomials. Integral \eqref{Pcoh1} then corresponds to a Weierstrass transform of that polynomial and is given by a remarkably simple expression
\begin{equation}\label{Imicr1}
J(x)=\frac 1{\sqrt{n!}} \left(\frac x{2\sigma}\right)^ne^{-x^2/8\sigma^2}.
\end{equation}
We see that, for objects of size $\lesssim\sigma$, photocurrent $J_{0n}$ gives the $n$th moment of the field in the object plane. 

The set of photocurrents acquired for multiple modes can be further utilized to find the decomposition of $E(x)$ into the Hermite-Gaussian basis and thereby reconstruct the full image of the object with a sub-Rayleigh resolution. Let $\alpha_{kn}$ be the coefficients of the Hermite polynomial of degree $k$, so that $H_k(x/{2\sigma})=\sum_n\alpha_{kn}(x/{2\sigma})^n$. Then, according to Eqs.~\eqref{Imicr} and \eqref{Imicr1}, we have
\begin{equation}\label{HGdecomp}
\beta_k:=\sum_n\sqrt{n!}\alpha_{kn}J_{on}=\intinf E(x)H_k\left(\frac x{2\sigma}\right)e^{-x^2/8\sigma^2}.
\end{equation}
Because Hermite-Gaussian functions form an orthonormal basis in the Hilbert space of one-dimensional functions, it follows that $E(x)=\sum_{k=0}^\infty\beta_kH_k(x/{2\sigma})e^{-x^2/8\sigma^2}$. Knowing all $\beta_k$, we can calculate $E(x)$. This approach is reminiscent of representing a quantum state of a harmonic oscillator as a superposition of Fock states, whose wavefunctions in the position basis are given by Hermite-Gaussian functions. 

We name the above-described imaging technique Hermite-Gaussian microscopy, or HGM. Acquiring heterodyne photocurrents for a sufficiently high number of TEMs in principle allows HGM to reconstruct the image with arbitrarily high resolution. The first few tens of TEMs, which are attainable in experimental optics, permit significant improvement of imaging with respect to the Rayleigh limit, as evidenced by Fig.~\ref{microscopyFig}.

The above conceptual description can be readily extended to practically relevant cases. Two-dimensional imaging is possible by scanning over both indices of TEM$_{mn}$ and measuring the photocurrent $J_{mn}$ for each pair $(m,n)$ up to a desired maximum. Acquiring both the sine and cosine quadratures of the heterodyne photocurrent permits phase-sensitive reconstruction of the object's field. 

A somewhat less trivial extension is to incoherent images. In this case, the output power of the heterodyne detector is given by
\begin{equation}\label{micrinc}
\avg{P}\propto\frac 1{n!}\intinf I(x)\left(\frac x{2\sigma}\right)^{2n}e^{-x^2/4\sigma^2}
\end{equation}
(see \emph{Supplementary} for the derivation). Similarly to the coherent case, we obtain moments of the field distribution in the object plane. However, now 
%the information is restricted to even moments only. Accordingly, 
we obtain only the even coefficients of the decomposition of $I(x)$ into the Hermite-Gaussian basis. Therefore the information about only the even component of function $I(x)$ is retained. An image reconstructed with these data will be a sum of the original intensity profile $I(x)$ with the ``ghost" image $I(-x)$. For two-dimensional microscopy, three ghost images, $I(x,-y)$, $I(-x,y)$ and $I(-x,-y)$, will be added to the original image $I(x,y)$. Their effect can be eliminated by placing the entire object into a single quadrant of the $x$-$y$ plane.

\paragraph{Summary.} We have used heterodyne detection in the TEM$_{01}$ Hermite-Gaussian mode to obtain sub-Rayleigh precision in the measurement of the position of single coherent and incoherent sources, as well as that of the separation between two incoherent sources. With numerical apertures below $10^{-3}$, our measurement precision is on a scale of a few microns. If our technique is used with state-of-the art microscopes with ${\rm NA}\sim 1$, precision on nanometer scales can be expected. By utilizing higher Hermite-Gaussian modes, the technique can be extended to full imaging with sub-Rayleigh resolution.

\paragraph{Note added.} While working on this manuscript, we became aware of similar research being pursued by Sheng {\it at al.} \cite{Ling}  and Tham {\it et al.} \cite{Tham}.

\paragraph{Acknowledgments} This work was supported by NSERC and CIFAR. We thank P. Barclay, R. Ghobadi, J. Moncreiff, A. Steinberg and M. Tsang for illuminating discussions, as well as B. Chawla for helping with the experiment.

\newpage\pagebreak\clearpage
\makeatletter
\renewcommand{\thefigure}{S\@arabic\c@figure}
\renewcommand{\theequation}{S\@arabic\c@equation}
\makeatother
\setcounter{figure}{0}

\setcounter{equation}{0}

\section{Supplementary Information}
\paragraph{Theoretical prediction for the signal.} Here we calculate the heterodyne detector signal in response to the electromagnetic field generated by an object of a specific shape. We used this method to obtain the theoretical curves in Figs.~2 and 3 of the main text. The calculation is for one-dimensional objects, but is readily extended to two dimensions.
 
We start by briefly reviewing Abbe's microscope resolution theory.% \cite{Hecht}. The object with spatial field distribution $E(x)$ generates the far-field distribution 
\begin{equation}\label{}
\tilde E(k_\perp)=\intinf E(x)e^{ik_\perp x}\de x,
\end{equation}
where $k_\perp$ is the orthogonal component of the wavevector and constant normalization factors are neglected throughout the calculation. The objective lens is located in the far field at distance $L$ from the object plane. The position $X$ in the lens plane is then related to  $k_\perp$ according to 
\begin{equation}\label{k2X}
X=L\frac{k_\perp}k=\frac{L\lambda k_\perp}{2\pi}.
\end{equation}
The lens, in turn, generates the inverse Fourier image in its focal plane:
\begin{align}\label{}
E'(x')&=\intinf \tilde E(k_\perp) \tilde T(k_\perp) e^{-ik_\perp x'}\de k_\perp\\
&=\intinf\intinf E(x) \tilde T(k_\perp) e^{ik_\perp (x-x')}\de x\de k_\perp,\nonumber
\end{align}
where $ \tilde T(k_\perp)$ is the transmissivity of the lens as a function of the transverse position in its plane. If this transmissivity is a constant, corresponding to an infinitely wide lens, the image is identical to the object: $E'(x')=E(x)$. If the lens is of finite width, the image is distorted according to 
\begin{equation}\label{Eprimeconv}
E'(x')=\intinf E(x) T(x'-x) \de x,
\end{equation}
where $T(x'-x)=\intinf \tilde T(k_\perp) e^{ik_\perp (x-x')}\de k_\perp$ is the Fourier image of the lens. In other words, the image is a convolution of the object with $T(\cdot)$. 

Heterodyne detection of the image field yields the electronic signal given by \eeqref{Pcoh} in the main text. If the field $E(x)$ is coherent, that equation is sufficient to calculate the output signal.

If the object is incoherent, we must take the average of the signal over all realizations of $E(x)$ to find the output power of the heterodyne detector photocurrent:
\begin{align}
\avg{P}&=\avg{\left(\intinf E'(x')E_{LO}(x')\de x'\right)^2}\\
&=\iintinf \avg{E'(x') E'(x'')}E_{LO}(x')E_{LO}(x'')\de x'\de x''\nonumber
\end{align}
Now using \eeqref{Eprimeconv} we find 
\begin{align}
&\avg{E'(x') E'(x'')}
\\&=\iintinf\avg{E(x_1) E(x_2)}T(x'-x_1) T(x''-x_2)\de x_1\de x_2.\nonumber
\end{align}
For an incoherent image, 
\begin{equation}\label{Idelta}
\avg{E(x_1) E(x_2)}=I(x_1)\delta(x_1-x_2)
\end{equation}
and hence 
\begin{align}\label{Pincoh}
\avg{P}&=\iiintinf I(x)T(x'-x) T(x''-x)\\
&\times E_{LO}(x')E_{LO}(x'')\de x\de x'\de x''.
\end{align}

Our goal is to determine the heterodyne detector signal as a function of the object configuration, $E(x)$ in the coherent case and $I(x)$ in the incoherent case. These calculations are simplified if we take into account a specific shape of the transmissivity function associated with a round diaphragm: $T(k_\perp)=\theta(k_{\perp{\rm max}}-|k|)$ with $k_{\perp{\rm max}}=2\pi R/(L\lambda)$ according to \eeqref{k2X}, $R=0.4\pm0.05$ mm is the radius of the diaphragm and $\theta(\cdot)$ is the Heaviside step function. In the Fourier domain this translates into 
\begin{equation}\label{}
T(x'-x)=\frac{J_1(k_{\perp{\rm max}}(x-x'))}{x-x'}\approx e^{-(x-x')^2/4\sigma^2},
\end{equation} where $J_1(\cdot)$ is a Bessel function. This is further approximated as $T(x'-x)\approx e^{-(x-x')^2/4\sigma^2}$ with $\sigma\approx0.21\lambda L/R=0.31\pm0.03$ mm.% \cite{Zhang07}. %The best fits between theory and experiments are however obtained for $\sigma=0.27$ mm.

The LO field in the TEM$_{00}$ mode is optimized to match the mode $E(x')$ in the coherent case for $E(x)=\delta(x)$, so $E_{LO}(x')=e^{-x'^2/4\sigma^2}$. Subsequently it is switched to the TEM$_{01}$ mode with $E_{LO}(x')=x'e^{-x'^2/4\sigma^2}$. Substituting this mode into the expression \eqref{Pcoh} in the main text for the coherent case, we find
\begin{equation}\label{}
P=J^2=\left(\intinf xE(x)e^{-x^2/8\sigma^2}\right)^2\de x.
\end{equation}
For the incoherent case, the mean signal power \eqref{Pincoh} equals
\begin{equation}\label{}
\avg{P}=\intinf x^2I(x)e^{-x^2/4\sigma^2}.
\end{equation}

\begin{figure}[h]
	\includegraphics[width=\columnwidth]{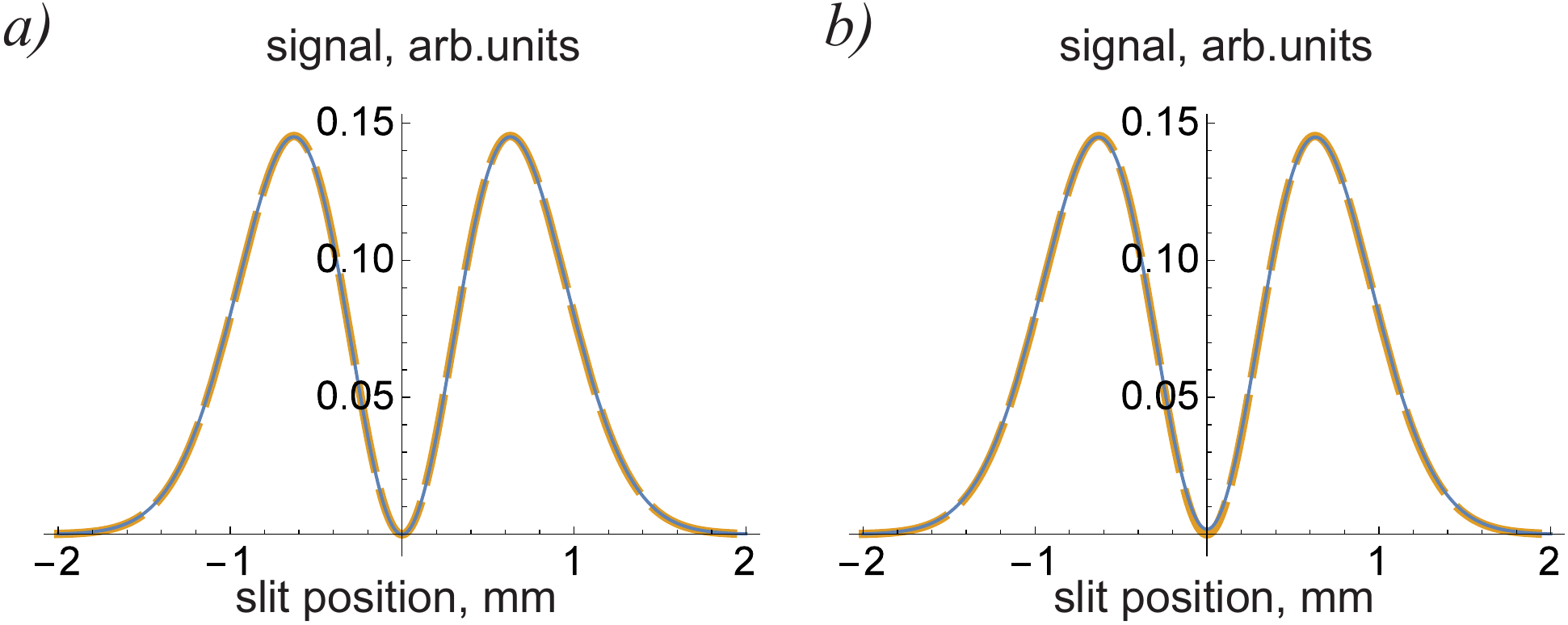}
	\caption{\label{compmodelsFig}Comparison of the theoretical predictions for the signal power taking into account the finite width of the slit (blue thin solid line) and assuming an infinitely narrow slit (yellow thick dashed line). a) Coherent case, b) incoherent case.}
\end{figure}

While the above results are valid for coherent and incoherent objects of any shape, a simple analysis leading to \eeqref{Pxp} in the main text is sufficient for the purposes of our experiment, as evidenced by Fig.~\ref{compmodelsFig}. The only visible difference between the two models is that the curve calculated for the incoherent case using the complete model does not reach zero at the slit position $x_p=0$. This is because an incoherent slit of a finite width, which can be seen as a combination of multiple mutually incoherent point sources positioned around $x_p=0$, makes a nonzero contribution to TEM$_{01}$.

\paragraph{Error analysis.} Our experimental setup yields the electronic signal power $P(x_p)$ with root mean square (rms) uncertainty $\sigma (P(x_p))$ as a function of the slit position $x_p$. Suppose the task is to estimate $x_p$ from the observed power. Below, we present a method for determining and minimizing the error of this estimation.

We can treat the observed power as a random variable whose probability distribution 
\begin{equation}\label{}
f(P,x_p)=\frac1{\sqrt{2\pi}\sigma}e^{-[P-P(x_p)]^2/2\sigma^2}
\end{equation}
is Gaussian with rms width $\sigma$ centered on the mean power $P(x_p)$. The problem then reduces to that of estimating parameter $x_p$ from this random variable. The uncertainty of this estimation can be found using the Cram\'er-Rao error bound,% \cite{ParamEst}, 
$\delta x_p\ge\sqrt{1/F}$, where 
$$F=\intinf \left[\frac{\partial f(P,x_p)}{\partial x_p}\right]^2\bigg{/}f(P,x_p)\de P$$
is the Fisher information. In the neighborhood of $x_p=0$, the power $P(x_p)$ can be approximated as $ax_p^2$, while the uncertainty $\sigma^2=c^2+[g P(x_p)]^2$ as discussed in the main text, with constants $a$, $c$ and $g$ evaluated from the experimental data. Accordingly, we find for the Fisher information in the limit $g\ll 1$
$$F=\frac{4 a^2 \theta ^2}{a^2 g^2 \theta ^4+c^2}.$$

Next, we determine the value of $x_p$ where the Fisher information is maximized so the measurement is the most sensitive. Taking the derivative of $F$, we find that the maximum value $F_{\max}=2a/gc$ is reached at $x_p=\sqrt{c/ag}$. It is these  values that we use to evaluate the estimation uncertainties of $x_p$ in the main text.

\paragraph{Derivation of \eeqref{micrinc} in the main text.}
We use \eeqref{Idelta} to write the output power of the heterodyne detector as the average 
\begin{align}\label{micrinc1}
\avg{P}&\propto\avg{J_{0n}^2}=\iintinf \avg{E(x_1) E(x_2)}J(x_1)J(x_2)\de x_1\de x_2\nonumber\\
&=\intinf I(x)J^2(x)\de x.
\end{align}
Substituting \eeqref{Imicr1} from the main text into the above result, we obtain \eeqref{micrinc}.

\end{document}